\RequirePackage{fix-cm}
\documentclass{svjour3}                     
\smartqed  
\usepackage{graphicx}
\usepackage{amssymb}
\usepackage{amsmath}
 \usepackage{mathptmx}      
\begin{document}

\title{Gravitational reduction of the wave function based on Bohmian quantum potential }

\author{Faramarz Rahmani\and Mehdi Golshani\and Ghadir Jafari}
\institute{F. Rahmani \at
              School of Physics, Institute for Research in Fundamental Science(IPM), Tehran, Iran\\
              Tel.: +98-21-22180692,
              Fax: +98-21-22280415\\
              \email{faramarzrahmani@ipm.ir}           
           \and
           M. Golshani \at
              School of Physics, Institute for Research in Fundamental Science(IPM), Tehran, Iran\\
              Department of Physics, Sharif University of Technology, Tehran, Iran\\
              Tel.:+98-21-66022718, Fax.:+98-21-66022718\\
              \email{golshani@sharif.edu}
\and
Gh. Jafari \at
              School of Physics, Institute for Research in Fundamental Science(IPM), Tehran, Iran\\
              Tel.: +98-21-22180692,
              Fax: +98-21-22280415\\
              \email{ghjafari@ipm.ir} }              
\date{Received: date / Accepted: date}

\maketitle

\begin{abstract}
In objective gravitational reduction of the wave function of a quantum system, the classical limit of the system is obtained in terms of the objective properties of the system. On the other hand, in Bohmian quantum mechanics the usual criterion for getting classical limit is the vanishing of the quantum potential or the quantum force of the system, which suffers from the lack of an objective description. In this regard, we investigated the usual criterion of getting the classical limit of a free particle in Bohmian quantum mechanics. Then we argued that how it is possible to have an objective gravitational classical limit related to the Bohmian mechanical concepts like quantum potential or quantum force. Also we derived a differential equation related to the wave function reduction. An interesting connection will be made between Bohmian concepts and gravitational concepts.
\end{abstract}

\section{Introduction}
\label{intro}
One of the mysterious issues in quantum mechanics is the wave function collapse or the wave function reduction. It has been studied in various contexts like the collapse hypothesis in standard or conventional quantum mechanics, many worlds interpretation, decoherence approach, gravitational reduction of the wave function and etc.\cite{Refsak,RefWh,RefG,RefV,RefBe,RefZ,RefMa,RefDe,RefDe2}. According to the conventional or standard quantum mechanics, the evolution of a quantum system is described by the Schr\"{o}dinger equation. It is a deterministic or unitary or probability preserving evolution. The word "deterministic" arises  from the fact that if we have the wave function at the initial time $t$, then it will be determined at the latter times by using the Schr\"{o}dinger equation:
\begin{equation}\label{schro}
i\hbar \frac{\partial \psi(\mathbf{x},t)}{\partial t}=\left(-\frac{\hbar^2}{2m}\nabla^2 + V(\mathbf{x},t) \right)\psi(\mathbf{x},t)
\end{equation}
where $V(\mathbf{x},t)$ represents an external potential.\par 
The linearity of the Schr\"{o}dinger equation allows us to consider the superposition of some solutions as a new solution. But this causes a strange behavior or feature for a quantum system. Because such a superposed state should describe the dynamics of a particle or body, while we have not seen that in the classical world a particle to be in a superposed state. In quantum mechanics, the usual justification for the superposition principle is that the particle is simultaneously in the all states before any observation. For example, before observation for detecting the direction of the spin of an electron, the electron is in a superposition of states up $\vert \uparrow \rangle$ and down $\vert \downarrow \rangle$ simultaneously. Another example is the famous Schr\"{o}dinger's cat which is alive or dead simultaneously before an observation.\cite{RefSch}. Only after observation we can talk about the death or life of the cat with certainty. A familiar example for the superposition of states, is a wave packet representing a free particle. In this case we assume that the particle is most probably in the spatial interval $\Delta x = \sigma$ where, $\sigma$ is the width of the wave packet. It is clear that this contradicts our daily experiences. In our daily experiences of the classical world, we do not observe objects in a superposition of different states simultaneously. Here, we can have at least two proposals. The first one is that the macroscopic world does not obey the quantum rules. Then, quantum mechanics is not universal and does not include the classical world. The second choice is that we assume that the quantum mechanics is universal and tends to the classical world continuously. In the latter case, the quantum and classical worlds should be seen as an undivided whole. The spirit of Bohmian quantum physics and the concept of quantum potential belongs to this deep view.\cite{RefBohm,RefHolland,RefImplicate,Refuniverse}. It is better to note beforehand that in Bohmian quantum mechanics the criterion for having classical limit is the vanishing of quantum potential or quantum force. We shall see how these quantities are involved in an objective model.\par
In general, in both Bohmian quantum mechanics and standard quantum mechanics, the process of wave function reduction is along with a measurement operation. But in Bohmian mechanics  the wave function reduction is an objective process in the sense that it does not need an observer to measure the specified quantity. Also, it resolves the non-unitary collapse of the wave function, by using the concept of empty waves and the possibility of definition of trajectories for the system and apparatus. The concept of empty waves will be clear latter.\par
The objective gravitational reduction of the wave function is based on the Penrose gravitational considerations.\cite{RefP1,RefP2,RefP3,RefP4}. In objective gravitational reduction, the existence of an observer is not necessary. In fact, Penrose has proven that the collapse hypothesis of standard quantum mechanics is understood through the interposition of gravitational effects. In a measurement processes, an apparatus is entangled with the quantum system. The apparatus is usually a macroscopic object. On the other hand, if we accept that quantum mechanics is universal, it should include the classical world and macroscopic objects. In fact, the apparatus (macroscopic object) is in a superposition of different states, but its self gravity reduces its quantum state vector to a specific state, and since the apparatus is entangled with the quantum system, the quantum system should also reduce to a specific state. This is the meaning of reduction in an objective gravitational reduction. In the other words, for the same reason we do not see the Schr\"{o}dinger cat or a macroscopic object in a superposed state, we do not see an electron in a superposition of different states simultaneously.\cite{RefP1,RefP2,RefP3}. Also, we expect that a macroscopic body  behaves classically. By using this idea and the considerations of objective gravitational reduction, a criterion for the mass of the particle or the body which is necessary for transition from quantum domain to classical domain is obtained. Before Penrose, Diosi has derived a relation for the minimal width of a wave packet in terms of the needed mass for which the particle or body behaves classically.\cite{RefD1}. It is based on the Schr\"{o}dinger-Newton equation. In that work the problem is to obtain an objective condition for the classical limit of a body. On the other hand, in the Penrose ideas for resolving the collapse hypothesis, the classical limit of a system is obtained objectively and gives the results of Diosi with more accuracy. Thus, the problem of wave function collapse and the classical limit of a quantum system have been resolved using the gravitational effects objectively. This, motivates us to investigate the classical limit of a free particle in Bohmian quantum mechanics and to see how is it possible to relate Bohmian mechanical concepts like quantum potential and quantum force, to the classical limit of a system objectively.?\par
Bohmian quantum physics is a deterministic and causal quantum theory which gives the same results as those of conventional quantum mechanics in experiments. The word "deterministic" in Bohmian quantum mechanics has a wider range relative to the conventional quantum mechanics. Because, in Bohmian quantum mechanics a quantum system is composed of a material system with physical properties that have been attributed to it like in classical mechanics with this difference that the dynamical quantities are obtained using the wave function of the system. In this theory, the wave function originates from a real agent which is not still clear to us. The wave function is represented in the configuration space and guides particles on some trajectories with definable positions and momentum, even before experiment or observation. Bohm's work is not a recovery of classical mechanics. Because in Bohm's causal theory an important quantity known as quantum potential is responsible for quantum motion of matter with non-classical features. Quantum potential and consequently quantum force act on a system in such a way that the system reaches the areas in the configuration space that are not accessible in classical mechanics.\cite{RefHolland}. The wave function in Bohmian quantum mechanics is not only a probabilistic tool; rather, its main task is to guide the quantum system causally.\cite{RefHolland}. In the other words, in Bohmian quantum mechanics, the probabilistic nature of quantum world is not intrinsic. Therefore, a relation like $\rho=\psi^{*}\psi$, is due to our ignorance of hidden variables that give quantum mechanical features to a system.\cite{RefHolland}. \par
In non-relativistic Bohmian quantum mechanics, the quantum motion of a particle is described by Schr\"{o}dinger's equation and the associated Hamilton-Jacobi equation. By writing the wave function in the polar form $\psi(\mathbf{x},t)=R(\mathbf{x},t)\exp(i\frac{S(\mathbf{x},t)}{\hbar})$ and substituting it into Schr\"{o}dinger's equation, we obtain the quantum Hamilton-Jacobi and continuity equations:
\begin{equation}\label{hamilton}
\frac{\partial S(\mathbf{x},t)}{\partial t}+\frac{(\nabla S)^2}{2m}+V(\mathbf{x})+Q(\mathbf{x})=0
\end{equation}
and 
\begin{equation}\label{con}
\frac{\partial R^2}{\partial t}+\frac{1}{m}\nabla \cdot(R^2 \nabla S)=0
\end{equation}
 The position of the particle is obtained from the following equation:
\begin{equation}\label{guidance}
\frac{d\mathbf{x}(t)}{dt}=\left(\frac{\nabla S(\mathbf{x},t)}{m}\right)_{\mathbf{X}=\mathbf{x}(t)}
\end{equation}
where $\nabla S(x,t)$ is the momentum of the particle and $ \rho=\psi^{*}\psi =R^2$. By knowing the initial position $\mathbf{x}_0$ and wave function $\psi(\mathbf{x}_0,t_0)$, the future of the system is obtained.The expression $\mathbf{X}=\mathbf{x}(t)$ means that among all possible trajectories, in an ensemble of particles, one of them is chosen. The quantity Q  in (\ref{hamilton}) is called quantum potential, and it is given by: 
\begin{equation}\label{potential}
Q=-\frac{\hbar^2 \nabla^2 R(\vec{x},t)}{2mR(\vec{x},t)}
\end{equation}
It may be said that the method by which  we get quantum Hamilton-Jacobi equation in Bohmian mechanics is somewhat ad hoc. But here we note that the substitution of the polar form of the wave function into the Schr\"{o}dinger equation is not the only approach towards Bohmian mechanics. The Hamilton-Jacobi equation and quantum potential is derivable also from another approach, without using the wave function and Schr\"{o}dinger's equation. Furthermore, for study a quantum system the usage of the Schr\"{o}dinger equation is not necessary and the equations (\ref{hamilton}) and (\ref{con}) are adequate.\cite{RefAtigh,RefAtigh2}.\par
In the following, we study the wave function reduction in conventional and Bohmian quantum mechanics briefly. Then we do a short review on the Penrose ideas about the wave function reduction.
After that we shall study the classical limit of a free particle in the context of Bohmian quantum mechanics through the concepts of quantum potential and quantum force. Then, we shall argue that the existence of a gravitational self interaction is necessary for having objective classical limit in Bohmian mechanics. The result that we obtain for the minimal width of an stationary wave packet for getting the classical limit, is the same that of the Diosi, which was obtained through the Schr\"{o}dinger-Newton equation. \cite{RefD1} But here, we shall derive it trough the concepts of Bohmian quantum mechanics. Finally, we shall get a nonlinear differential equation for mass distribution at the classical limit.
\section{Constructing an objective classical limit in the Bohmian context}
\label{sec:1}
In the conventional or standard quantum mechanics, all information that we need to describe a quantum system exists in the wave function of that system. The wave function does not point to any reality. It is only a probabilistic instrument which is interpreted as the knowledge of observer or experimenter about the system.\cite{RefP4}. In fact, our knowledge about a physical system is summarized in the prediction of the probability of measurement of a specific eigenvalue for a specific physical quantity, like energy, momentum, spin direction, etc. It is clear that this not an ontological view. This is known as Copenhagen interpretation of quantum mechanics. By the evolution of a quantum system, we mean that 
the probabilistic wave $\psi$ of a system has unitary evolution governed by the Schr\"{o}dinger equation. In this context, the evolution of the physical system is deterministic. Because, having the wave function at an initial time, the Schr\"{o}dinger equation gives its evolution at the latter times. \par 
According to the postulates of standard quantum mechanics, the measurement operation collapses the state vector $\vert \psi \rangle$ of a system to one of its eigenvectors instantaneously. This is not a unitary evolution, because it takes place instantaneously and also the process of measurement is a random jumping from  a continuous evolution process to a mixture of some states. In other words:
\begin{displaymath}
\psi=\left(\sum_i a_i \psi_i \right)\otimes \phi_0  \xrightarrow[\text{measurement}]{\text{random jump}} \psi_i \otimes \phi_i, \quad \text{with the detection probability}\quad \vert a_i \vert ^2 
\end{displaymath}
where, $\psi$ is the total wave function of the system plus apparatus. The initial state of the apparatus is $\phi_0$ and its state after measurement is $\phi_i$. This is an ideal measurement in which the system state $\sum_i a_i \psi_i $ does not alter during the measurement process.\cite{RefHolland}
In a specific measurement we can not predict which of the eigenvalues will be detected. The above statistical jump does not conserves the unitarity condition. See ref \cite{RefHolland}.
In the conventional quantum mechanics, the measurement apparatus is a classical object. This means that there is sharp distinction between classical and quantum world in the conventional quantum mechanics.\par  
In Bohmian quantum mechanics, the situation is somewhat better. There is no need to an observer to register an eigenvalue of the quantum system. There, the measurement apparatus and quantum system are interacting and there is an interaction Hamiltonian which participate in the total dynamics of the system and apparatus like in standard quantum mechanics. But, in Bohmian quantum mechanics, a quantum system consists of a pilot wave and the particle(s) or body with definable trajectories. Also, the apparatus obeys the quantum rules. Thus, in a position measurement the pointer of the apparatus which has a definite trajectory will be determined using the total state of the system automatically without needing an observer. Thus, the need for an observer is removed in this interpretation. The collapse hypothesis is also removed, but with another novel concept namely "empty waves" comes in. In Bohmian quantum mechanics, a particle chooses one of the trajectories among the possible trajectories of the system . In this situation, the associated wave function is not empty but other possible states are empty. Also, the apparatus lies in a specified state with its associated trajectory, whether an observer is present or not. The other states remain empty and go away after measurement. So, the collapse hypothesis is not necessary. \par It is noteworthy the empty waves affect the dynamics of the system through the superposition of all possible states. For example in the two-slit experiment, when a wave $\psi$, splits into the two packets $\psi_1$ and $ \psi_2$, the particle is in one of the traversed routs, not in both of them simultaneously. Because the single-valuedness of the wave function does not allow the trajectories cross the each other.\cite{RefHolland}. Hence, when the particle is in one of the routs, the other wave, for example, $\psi_2$ is an empty wave and vice-versa. But the important point is that the empty waves affect the dynamics of the system through the interference of waves. The usefulness of the empty packets is that it avoids the collapse hypothesis. More clearly, when a measurement takes place the total wave function somehow leads to the state in which the particle is present and other empty waves go their own way. This demonstrates that in Bohmian quantum mechanics there is no sharp distinction between classical and quantum world. For details see, \cite{RefHolland}. This type of interaction does not contradict the  unitary evolution of the Schr\"{o}dinger equation in the presence of measurements. Because, the process goes on continuously without the losing of information by the aid of empty waves. \par 
Here there are two problems. First, is the existence of empty waves and the problem of detecting them. The second, is that it is not an objective reduction in the sense that we can determine the reduction time or the needed reduction mass for a particle or body. We know that the state vector of an electron is reduced through a measurement process. But how our universe as a macroscopic body is reduced?,  and through which measurement or apparatus?  Thus, we look for an internal agent for reducing the system whether the system is a sub atomic particle or the whole of universe. It seems that by increasing the mass of a particle or system, its dynamics tends to classical deterministic dynamics. In the other words, a classical object is in a localized state not in a superposed one. But how does this occur systematically? Penrose has proven this fact through the self-gravitational effects of the particle or body. Before him some authors like Karolyhazy and Diosi had studied the relation between the self-gravity of a body and the classical limit of the system. \cite{RefD1,RefK,RefD2,RefD3}. 
As we mentioned before, resolving the collapse problem through the gravitational considerations, also gives the classical limit of the system. In other words, due to the effects of gravity, a mechanism is set up that relates the classical limit of a quantum system and its wave function reduction objectively. Therefore, through the properties of the particle the classical limit is determined. This motivates us to use the concept of gravitational reduction in a Bohmian context to make a clear objective criterion for having classical limit related to quantum potential or quantum force.\par 
Penrose has two viewpoints on the reduction of wave function in refs \cite{RefP1} and \cite{RefP3}, based on the equivalence principle of general relativity and the principle of general covariance. In the first approach, the principle of equivalence leads to a phase difference in the wave function of a falling body with respect to the wave function of body that experiences gravitational force. This difference generates a new term in the energy of the body that measures the uncertainty in the gravitational self energy of the particle and gives an objective criterion for transferring from quantum phase to the classical phase. In the latter case, the principle of general covariance leads us to conclude that considering the quantum states of the self gravity of a body leads to different vacuums with different Killing vectors, like what happens in the Unruh effect. But here, the problem is studied non-relativistically. Here, we look at the general themes of Penrose's ideas briefly.\par 
Consider a non-relativistic falling body in a constant gravitational field $\mathbf{g}$ with the coordinate system $(\mathbf{x},t)$. The Schro\"{o}dinger equation for this case becomes:
\begin{equation}\label{ns}
i\hbar \frac{\partial\psi}{\partial t}=-\frac{\hbar^2}{2m}\nabla^2 \psi -m\mathbf{x}\cdot \mathbf{g}\psi
\end{equation}
According to the equivalence principle, this is also a freely falling particle. If we choose the coordinate $(\mathbf{X},T=t)$ for freely falling motion, the Schr\"{o}dinger equation for it becomes:
\begin{equation}\label{rs}
i\hbar \frac{\partial\Psi}{\partial T}=-\frac{\hbar^2}{2m}\nabla^2 \Psi
\end{equation}
Where, $\Psi$ is the free falling wave function. These two coordinates are related as $\mathbf{x}=\mathbf{X}+\frac{1}{2}\mathbf{g}t^2$. Also, we have $\nabla^2_{\mathbf{X}}=\nabla_{\mathbf{x}}^2=\nabla^2$. For establishing the equivalence principle, the two wave functions $\psi$ and $\Psi$ should be related as:
\begin{equation}\label{wr}
\Psi=\exp\left(\frac{im}{\hbar}\left(\frac{g^2 t^3}{6}-\mathbf{x}\cdot \mathbf{g} t\right)\right)\psi
\end{equation}
or
\begin{equation}\label{wr2}
\psi=\exp\left(\frac{im}{\hbar}\left(\frac{g^2 T^3}{3}+\mathbf{X}\cdot \mathbf{g} T\right)\right)\Psi
\end{equation}
In ref \cite{RefP3} the important role of the term $\frac{im g^2 t^3}{6\hbar} $ has been explained. This term, refers to a new vacuum and the issue of the Unruh effect, in which a pure quantum state in an accelerated frame or in a gravitational field is seen as a mixture of different states. On the other hand, we know that the measurement process or reduction of the wave function leads to some statistical mixture of information about a quantum system. This means that both of them may have identical origin.\cite{RefP1,RefP2,RefP3}. \par 
The other consideration of Penrose is based on the concept of Killing vectors. If, we consider the quantum states of the self-gravity of the particle, when it is in two different locations, the superposed state does not include a unique time-like Killing vector. We know that for having a stationary spacetime the existence of a time-like Killing vector is necessary. In this case, such unique Killing vector is not definable. So the superposed state decays to a single state for which the definition of a time-like Killing vector is possible. Some conditions are obtained for the needed mass of the particle and the width of its associated wave packet for transition from quantum domain to classical domain through these arguments. This is an objective gravitational reduction description. Because, it is determined by the properties of the particle or body like its mass.\par
Some related results have also been obtained by Diosi from another point of view, based on the Schr\"{o}dinger-Newton equation. The  Schr\"{o}dinger-Newton equation describes the quantum dynamics of a system  that are affected by its self gravitational fields. For a single body this equation is:
\begin{equation}\label{sn}
i\hbar \frac{\partial\psi(\mathbf{x},t)}{\partial t}=\left(-\frac{\hbar^2}{2M}\nabla^2 -GM^2 \int \frac{\vert \psi(\mathbf{x}^\prime,t)\vert^2}{\vert \mathbf{x}^\prime -\mathbf{x} \vert} d^3 x^\prime\right) \psi(\mathbf{x},t)
\end{equation}
By using a stationary state $\psi=\psi_0 e^\frac{iEt}{\hbar}$, which satisfies the above equation, a relation is obtained between the mass of the particle and the width of its associated stationary wave packet which provides a criterion for the transition from the quantum world to the classical world.\cite{RefD1}. In the following, we first argue that the usual Bohmian condition of transition from quantum mechanics to classical mechanics for a wave packet is not suitable for an objective reduction. Then we investigate this problem by using the concepts of Bohmian quantum potential and Bohmian quantum force in such a way that leads us to an objective Bohmian reduction (classical limit).\par In summary, gravitation would localize a bulk  of matter. In the language of Bohmian quantum mechanics, it would localize the ensemble of different locations of a particle or body. In fact, the concept of gravitational localization should be generalized. A free particle is described by a spreading wave packet, according to the Schro\"{o}dinger equation. This dispersion is a quantum mechanical effect which is explained by the use of the Heisenberg uncertainty principle in conventional quantum mechanics. In Bohmian quantum mechanics, this is described with the concept of quantum force.\cite{RefHolland}\par 
In Bohmian quantum mechanics, the quantum potential has important properties. In Bohm's own view, it is responsible for the quantum motion of matter. In some other views, like that of DGZ, the quantum potential is not necessary to describe the dynamics of the system and the guidance equation (\ref{guidance}) is sufficient for determining the dynamics of the particle.\cite{RefGold}. But in the both of them the vanishing of quantum potential or quantum force ($f=-\nabla Q$) is a main criterion for the transition from quantum domain to classical domain. In conventional quantum mechanics, the condition for transition from quantum to classical domain is the vanishing of the Plank constant ($\hbar \longrightarrow 0$), which is not an acceptable condition. Its conflicts with Bohmian conditions for transition from quantum to classical world i.e, the vanishing of quantum potential or quantum force, is studied in ref \cite{RefHolland}. \par 
Now, we want to look at the issue through the study of the dynamics of wave packets in Bohmian quantum mechanics. A free wave packet, satisfying the Schr\"{o}dinger equation, is:
\begin{equation}\label{packet}
\psi(\mathbf{x},t)=(2\pi s_t^2)^{-\frac{3}{4}} e^{\left(i\mathbf{k}\cdot ( \mathbf{x}-\frac{\mathbf{u}t}{2}) -\frac{(\mathbf{x}-\mathbf{u}t)^2}{4s_t \sigma_0}\right) }
\end{equation}
Where, $\mathbf{u}=\frac{\hbar \mathbf{k}}{m}$ is the initial group velocity of the center of the packet.\cite{RefHolland}. The $\sigma_0$ is the random mean square width of the packet. The $s_t$ is defined as $s_t = \sigma_0(1+\frac{i\hbar t}{2m\sigma_0^2})^{\frac{1}{2}}$. The random mean square width of the packet at time $t$ is defined as 
\begin{equation}\label{w}
\sigma = \vert s_t \vert =\sigma_0 \left(1+ (\frac{\hbar t}{2m\sigma_0^2})^2\right)^{\frac{1}{2}}
\end{equation}
which represents the spreading of the wave packet. The amplitude and the phase of the packet are:
\begin{equation}\label{rf}
R= (2\pi \sigma^2)^{-\frac{3}{4}} e^{-\frac{(\mathbf{x}-\mathbf{u}t)^2}{4\sigma^2}}
\end{equation}
and
\begin{equation}\label{sf}
S = -(\frac{3\hbar}{2})\arctan(\frac{\hbar t}{2m\sigma_0^2})+m\mathbf{u}\cdot (\mathbf{x}-\frac{1}{2}\mathbf{u}t)+\frac{(\mathbf{x}-\mathbf{u}t)^2}{8m\sigma_0^2 \sigma^2}
\end{equation}
The quantum potential and quantum force for this system is obtained through the relations:
\begin{equation}\label{qf}
Q= \frac{\hbar^2}{4m\sigma^2}\left(3-\frac{(\mathbf{x}-\mathbf{u}t)^2}{2\sigma^2} \right)
\end{equation}
and
\begin{equation}\label{ff}
f=-\nabla Q = \frac{\hbar^2}{4m\sigma^2}(\mathbf{x}-\mathbf{u}t)
\end{equation}
Now we want to argue that the usual Bohmian condition for getting classical limit is not suitable from an objective point of view. As we mentioned before, in conventional quantum mechanics, the explanation for spreading of the wave packet is based on the Heisenberg uncertainty relation.
In Bohmian quantum mechanics, the spreading of wave packet is due to quantum force \cite{RefHolland}. The condition for the classical limit is the vanishing of the quantum force or quantum potential. When quantum potential vanishes the quantum Hamilton-Jacobi equation reduces to classical Hamilton-Jacobi equation. In Bohmian quantum mechanics, there is no explanation for the vanishing of quantum potential or quantum force objectively. The formalism only states that if quantum potential or force vanishes, the classical circumstances will be retrieved. \par 
Here, we investigate the  classical limit of a free particle with its associated wave packet (\ref{packet}).
The vanishing of the quantum potential or the quantum force of the wave packet is based on this argument that in classical limit the wave packet does not spread i.e. we have $\sigma \longrightarrow \sigma_0$. Thus, we should impose the condition
\begin{equation}\label{c1}
\frac{\hbar t}{2m\sigma_0^2} \longrightarrow 0
\end{equation}
on the relation (\ref{w}). By this condition, the amplitude and the phase of the wave packet become:
\begin{equation}
R \longrightarrow (2\pi \sigma_0^2)^{-\frac{3}{4}} e^{-\frac{(\mathbf{x}-\mathbf{u}t)^2}{4\sigma_0^2}}\\
\end{equation}
and
\begin{equation}
S \longrightarrow m\mathbf{u}\cdot \mathbf{x}- Et
\end{equation}
with the classical constant energy $E=\frac{1}{2}m \mathbf{u}^2$. The condition (\ref{c1}), states that if the initial width of the wave packet or the mass of the particle is very large, at the initial times, the fraction $\frac{\hbar t}{2m\sigma_0^2}$ is too small and the dispersion of the wave packet is negligible. From the experimental view this is a flawless condition. Because, before increasing the magnitude of the fraction (\ref{c1}) over time, we can do a measurement. But from an objective point of view, it is not a convincing condition. Suppose that we let the time in numerator of the relation (\ref{w}) grows to infinity. Naturally, the above condition fails. Because, the numerator and denominator will be comparable. But, we are interested in a condition that expresses the sufficient amount of mass to get classical limit, independent of time. Another defect is that, it is possible to have large mass in the relation (\ref{w}) with the width of the wave packet being very small. Since the width of the wave packet refers to wave function or uncontrollable hidden variables, the relation (\ref{w}) does not give an objective criterion. We should replace $\sigma$  with the properties of the object or universal constants which are measurable. We have seen experimentally that by increasing the mass of a particle its dynamics tends to classical dynamics. In the classical dynamics, the particle has a precise position, while in quantum mechanics there is an uncertainty in its position. We note that in the standard quantum mechanics the position of a particle is measured through the action of its associated operator on the wave function, while in Bohmian quantum mechanics the particle has specific position independent of operator formalism and measurement theory. \par
\begin{figure}[ht] 
\centerline{\includegraphics[width=6cm]{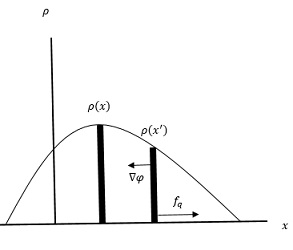}}
\caption{The quantum force of the wave packet spreads the wave packet and increases the uncertainty in the particle location. On the other hand the gravitation between mass distribution at different positions of the particle in the ensemble would localize the distribution and decreases the uncertainty in the position of the particle. The quantum force is depicted schematically.\label{fig:1}}
\end{figure}

As we mentioned earlier the quantum force is the responsible for the spreading of the wave packet. Dynamically, we need a force that prevents the quantum force to spread the wave function. Since, we do not think about external agents, we have to find this force in the particle's own properties. The best candidate among forces is the gravity which is always attractive and would localize the mass distribution. Note that it is the self gravity of a system which is important, not the gravitation due to other objects. Because, the gravitation due to other objects does not localize the different locations of a body in the ensemble. Here, the localization is a more general concept than the localization in classical mechanics. It should be applicable even for a point-like particle. In fact, we should have a quantum mechanical view about the effects of gravity. In the other words, we know from Bohmian quantum mechanics that for a particle at an initial time $t_0$ with the initial wave function $\psi_0$, there is an ensemble of positions and consequently trajectories which are distributed in space. So, we can consider the gravitational interaction between the different locations of particle in the ensemble. In figure \ref{fig:1}, we illustrate the gravitational force between different particle locations in a wave packet. In Bohmian mechanics, a particle can be at different points of a wave packet. In fact, since the hidden variable(s) are not clear to us, we think that the particle is in all possible locations of ensemble simultaneously. However, this behavior is justified with the uncertainty principle in standard quantum mechanics, but in Bohmian quantum mechanics such behaviors are due to our ignorance with respect to the nonlocal hidden variables\cite{RefHolland}.\par 
At the classical limit two statements are possible. First, the quantum force and self gravity of particles are equal. In this case, we have a stationary non-spreading wave packet with the constant width $\sigma_0$ as:
\begin{equation}
\psi \longrightarrow R_0(\mathbf{x}) exp(\frac{iEt}{\hbar})
\end{equation}
Second, the self gravity of the particle overcomes the quantum force completely and the wave packet tends to a Dirac delta function, i.e.
\begin{equation}
\psi \longrightarrow \delta^3(\mathbf{x}-\mathbf{x}^\prime)
\end{equation}
But, the final state can not be a Dirac delta function. Because, the width of the Dirac delta function tends to zero and this causes an infinite quantum force. It seems that an equivalence should be between quantum force and self gravitation of the particle at the classical limit.
In fact, according to the figure (\ref{fig:1}), the wave packet begins to disperse due to the quantum force. The relation (\ref{ff}) demonstrates that if initial width is more narrow, the force is  stronger and the the wave packet spreads rapidly. At the same time, the self gravitation of the particle (gravitation in ensemble) prevents more dispersion. If the gravity is so strong that overcomes the spreading of the wave function, then after a specific time the quantum dispersion vanishes. Then, we should have at the equivalence time: 
\begin{equation}\label{fe}
\mathbf{f}_q = \mathbf{f}_{\mathfrak{g}}
\end{equation}
The relation (\ref{fe}) is locally equal to:
\begin{equation}\label{fe2}
\nabla Q(\mathbf{x}) =m \nabla \varphi(\mathbf{x}) 
\end{equation}
Through this argument, we conclude that we should add a gravitational self interaction to the quantum Hamilton-Jacobi of the particle. Because, the dynamics of the particle is determined through the all potentials of the system i.e., $m\mathbf{a}= \nabla(Q+\sum_i U_i)$. For the two specified elements of the figure (\ref{fig:1}), the gravitational potential is
\begin{equation}
d \varphi (\mathbf{x}, \mathbf{x}^\prime)=-\frac{G m \rho(\mathbf{x}^\prime)d^3 \mathbf{x}^\prime}{\vert \mathbf{x}-\mathbf{x}^\prime \vert} 
\end{equation}
The gravitational energy for all possible particle locations is:
\begin{equation}
U_{\mathfrak{g}}(\mathbf{x})= -m^2 \int \frac{G \rho(\mathbf{x}^\prime)d^3 x^\prime}{\vert \mathbf{x}-\mathbf{x}^\prime \vert} 
\end{equation}
Note that in figure (\ref{fig:1}) the quantum force is depicted schematically. By the specified force in figure (\ref{fig:1}), we do not mean the quantum force as a repulsive force between the elements of the ensemble, like the forces of the classical mechanics. However, this force would make more uncertainty in the location of the particle.
Thus, the new quantum Hamilton-Jacobi equation becomes:
\begin{equation}\label{hg}
\frac{\partial S}{\partial t}+\frac{(\nabla S)^2}{2m}+ Q(\mathbf{x})- m^2G \int \frac{ \vert \psi (\mathbf{x}^\prime)\vert^2 }{\vert \mathbf{x}-\mathbf{x}^\prime \vert}d^3 \mathbf{x}^\prime=0
\end{equation}
Where, we have used $\rho (\mathbf{x}^\prime)=\vert \psi (\mathbf{x}^\prime)\vert^2 $. It is not difficult to check that the substitution the polar form of the wave function into the Schr\"{o}dinger-Newton equation leads to the above Hamilton-Jacobi equation. But, we suggested to get the above Hamilton-Jacobi equation through some special arguments.
The average of the above Hamilton-Jacobi equation is:
\begin{equation}
\int \rho(\mathbf{x})\left( \frac{\partial S}{\partial t}+\frac{(\nabla S)^2}{2m}+ Q(\mathbf{x})- m^2G \int \frac{ \vert \psi (\mathbf{x}^\prime)\vert^2 d^3 x^\prime}{\vert \mathbf{x}-\mathbf{x}^\prime \vert} \right)d^3\mathbf{x}=0
\end{equation}
Or equivalently,
\begin{equation}
\int  \left( -E+\frac{\mathbf{p}^2}{2m}+ Q(\mathbf{x})- m^2G \int \frac{ \vert \psi (\mathbf{x}^\prime)\vert^2 d^3 x^\prime}{\vert \mathbf{x}-\mathbf{x}^\prime \vert} \right) \vert \psi(\mathbf{x})\vert^2 d^3\mathbf{x}=0
\end{equation}
In an abbreviated form, we have:
\begin{equation}\label{abb}
\left\langle E \right\rangle = \left\langle \frac{\mathbf{p}^2}{2m} \right\rangle + \left\langle Q(\mathbf{x}) \right\rangle + \left\langle U_\mathfrak{g} \right\rangle
\end{equation}
For simplicity, we do calculations for a one-dimensional wave packet with the width $\sigma_0$ and zero initial group velocity.
If we calculate the quantum potential for a stationary one-dimensional wave packet $\psi(x,t)= (2\pi \sigma_0^2)^{-\frac{3}{4}}e^{-\frac{x^2}{4\sigma_0^2}} e^{\frac{iEt}{\hbar}}$, in which $R_0=(2\pi \sigma_0^2)^{-\frac{3}{4}}e^{-\frac{x^2}{4\sigma_0^2}} $, we get:
\begin{equation}\label{aq}
\langle Q\rangle _{0}=\int_{-\infty}^{+\infty} R_0^2  Q  dx  \sim \frac{\hbar^2}{2m\sigma_0^2}
\end{equation}
which is the average quantum potential of the particle, when it is described by a stationary wave packet with the width  $\sigma_0$. The average gravitational self energy of a point-like particle, with the probability radius $\sigma_0$, is 
\begin{equation}
\langle U _{\mathfrak{g}}\rangle = \int_{-\infty}^{+\infty} R_0^2 U _{\mathfrak{g}}   dx \sim \frac{Gm^2}{\sigma_0}
\end{equation}  
A macroscopic body with radius $R$ has been studied in refs \cite{RefP3} and \cite{RefD2}. For simplicity, we consider a point-like particle. Because our aim is only the study of gravitational reduction in Bohmian context. 
Since for a stationary state, the phase of the wave packet is independent of position we conclude that $p=\nabla S=0$. Thus the kinetic energy in the Hamilton-Jacobi equation is zero. Note that this statement is possible in Bohmian quantum mechanics and not in the standard quantum mechanics. Thus, the relation (\ref{abb}) becomes:
\begin{equation}\label{eg}
\langle E \rangle =\frac{\hbar^2}{2m\sigma_0^2}-\frac{Gm^2}{\sigma_0}
\end{equation}
For having equilibrium of forces as the average, we calculate $\frac{d \langle Q\rangle _{0}}{d\sigma_0}=\frac{d \langle U _{\mathfrak{g}}\rangle}{d \sigma_0}$, which yields condition
\begin{equation}\label{min}
(\sigma_0)_{\text{min}}\sim \frac{\hbar^2}{Gm^3}
\end{equation}
This is the famous result of ref. \cite{RefD1} which we have derived here through the Bohmian considerations. In this relation, the width of the wave packet which is related to a subjective concept, like hidden variable, is related to an objective quantity like the mass of the particle. By using the relation (\ref{min}), the average stationary quantum potential is represented by measurable quantities such as 
\begin{equation}
\langle Q\rangle _{(stationary)_{min}}\sim \frac{G^2 m^5}{\hbar^2}
\end{equation}
This is a criterion for the average quantum potential to give the classical limit for a free particle with mass $m$ objectively, independent of whether the particle is an electron or a macroscopic body.\par
If we operate with the operator $(\nabla \cdot) $ on the both sides of the relation (\ref{fe2}), we get:
\begin{equation}\label{con2}
\nabla^2 Q = 4\pi G m \rho 
\end{equation}
where, we have used the Poisson equation $\nabla^2 \varphi =4\pi G \rho $. This is  an interesting result. Because, it represents a nonlinear differential equation for the objective Bohmian classical limit. The relation (\ref{con2}), has an interesting interpretation. It states that at the classical limit, for which the quantum and gravitational forces are equal, the quantum information which is in general non-local, reduces to local gravitational information. The relation (\ref{con2}) can be represented as
\begin{equation}\label{ee}
\frac{\hbar^2}{2m^2}\nabla^2(\frac{\nabla^2 \sqrt{\rho}}{\sqrt{\rho}}) =-4\pi G  \rho
\end{equation}
which represents  a stationary quantum-gravitational bulk in space. It is possible to solve this equation analytically or numerically and investigate the solutions of $\rho$. \par
\begin{figure}[ht] 
\centerline{\includegraphics[width=9cm]{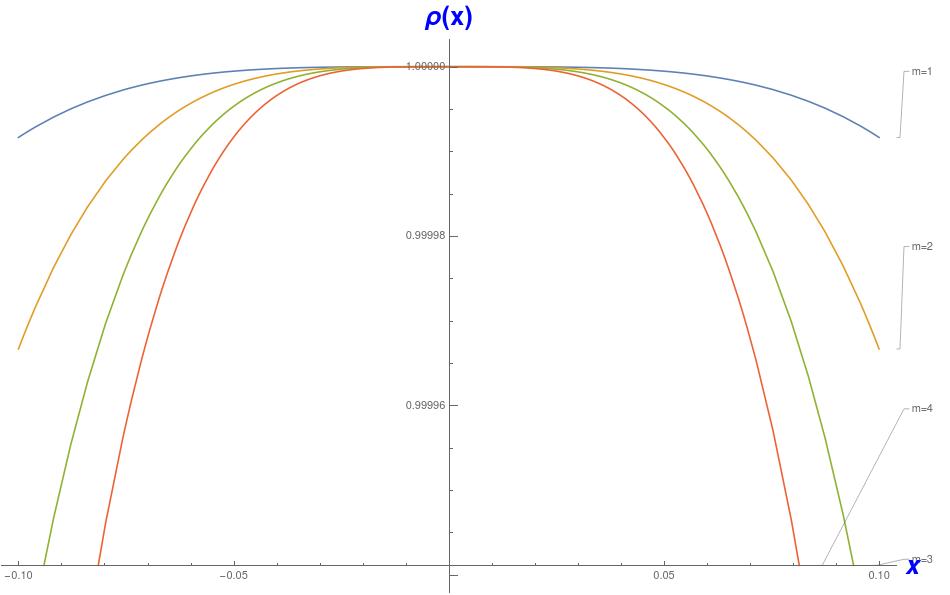}}
\caption{Different solutions of (\ref{ee}) for various values of $m$. 
The width of distribution of $\rho$ is proportional to $\sigma_0 \propto \frac{1}{\sqrt{m}}$ .\label{fig:4}}
\end{figure}

 
In the following, the figure (\ref{fig:4}) illustrates different solutions of the equation (\ref{ee}) for different values of mass. It demonstrates how the probability of mass distribution is concentrated by the increase in the mass of the particle. 
These arguments demonstrate that how an important quantity such as Bohmian quantum potential, which in Bohm's own view is the responsible for quantum behaviors of matter, is related to the gravitational potential, and gives an objective criterion for the classical limit of a free particle in Bohmian quantum mechanics. 
\section{conclusion}
\label{sec:3}
In this work, we demonstrated that how the famous result of the gravitational wave function reduction i.e. the relation (\ref{min}), is obtained through the considerations of Bohmian quantum mechanics. We derived an objective condition for the transition from the quantum domain to the classical domain, using the concept of Bohmian quantum mechanics. The practical criterion is the equivalence of the average quantum force and the average self gravitational force of a body. Also, we obtained a relation for the average of quantum potential, relation (\ref{aq}), in terms of measurable quantities like the mass of the particle. The study of wave function reduction in the context of Bohmian quantum mechanics leads to a quantum-gravitational matter bulk that its associated equation can be solved analytically or numerically for getting more understanding. In fact, we have demonstrated that quantum information  reduces to gravitational information at the reduction time. It represents a deep relation between quantum mechanics and gravity which should be studied more.
Also, in this approach, the non-objective classical limit in Bohmian quantum mechanics, i.e., the vanishing of quantum potential or quantum force, is modified somehow, and could be expressed in terms of objective parameters like the mass of the particle. Another achievement is that in this approach, the particle participates in its quantum state reduction through the its mass and its self gravity. But, in the usual Bohmian classical limit all that happens is the one-way effects of the wave function, and the mass and gravity of the particle has no direct role in obtaining classical limit of the system. Therefore,  contrary to usual Bohmian criterion, and the authors of the ref \cite{RefB}, the particle has active role in its quantum state reduction for obtaining objective Bohmian classical limit.



\end{document}